\documentclass[10pt, a4paper, conference]{IEEEtran}

\IEEEoverridecommandlockouts
\usepackage{cite}
\usepackage{url}
\usepackage{amsmath,amssymb,amsfonts}
\usepackage{algorithmic}
\usepackage{graphicx}
\usepackage{adjustbox}
\usepackage{textcomp}
\usepackage{multirow}
\usepackage{comment}
\usepackage[caption=false]{subfig}
\usepackage{xcolor}
\def\BibTeX{{\rm B\kern-.05em{\sc i\kern-.025em b}\kern-.08em
    T\kern-.1667em\lower.7ex\hbox{E}\kern-.125emX}}
\begin{document}

\title{AI-Driven Vehicle Condition Monitoring with Cell-Aware Edge Service Migration\\

\thanks{This work has been partially supported by the ``Ministerio de Asuntos Económicos y Transformación Digital" and the European Union-NextGenerationEU in the frameworks of the ``Plan de Recuperación, Transformación y Resiliencia" and of the ``Mecanismo de Recuperación y Resiliencia" under 
project SUCCESS-6G with reference TSI-063000-2021-39, TSI-063000-2021-40, and TSI-063000-2021-41. Pavol Mulinka performed this research while at the Centre Tecnològic de Telecomunicacions de Catalunya (CTTC/CERCA), Castelldefels, Spain.}
}

\author{\IEEEauthorblockN{Charalampos Kalalas\IEEEauthorrefmark{1},
		Pavol Mulinka\IEEEauthorrefmark{2}, Guillermo Candela Belmonte\IEEEauthorrefmark{3}, Miguel Fornell\IEEEauthorrefmark{4}, Michail Dalgitsis\IEEEauthorrefmark{5},\\ Francisco Paredes Vera\IEEEauthorrefmark{4}, Javier Santaella Sánchez\IEEEauthorrefmark{6}, Carmen Vicente Villares\IEEEauthorrefmark{6}, Roshan Sedar\IEEEauthorrefmark{1}, Eftychia Datsika\IEEEauthorrefmark{5},\\ Angelos Antonopoulos\IEEEauthorrefmark{5}, Antonio Fernández Ojea\IEEEauthorrefmark{3}, and
		Miquel Payaro\IEEEauthorrefmark{1}}
	\IEEEauthorblockA{\IEEEauthorrefmark{1}Centre Tecnològic de Telecomunicacions de Catalunya (CTTC/CERCA), Castelldefels, Spain\\}
	\IEEEauthorblockA{\IEEEauthorrefmark{2}Forescout Technologies, Eindhoven, Netherlands\\}
    \IEEEauthorblockA{\IEEEauthorrefmark{3}Optare Solutions, Vigo, Spain\\}
    \IEEEauthorblockA{\IEEEauthorrefmark{4}Idneo Technologies, Mollet del Vallès, Spain\\}
    \IEEEauthorblockA{\IEEEauthorrefmark{5}Nearby Computing, Barcelona, Spain\\}
    \IEEEauthorblockA{\IEEEauthorrefmark{6}Cellnex Telecom, Barcelona, Spain\\
	Email: \{ckalalas@cttc.es\}}
	}

\maketitle

\begin{abstract}
Artificial intelligence (AI) has been increasingly applied to the condition monitoring of vehicular equipment, aiming to enhance maintenance strategies, reduce costs, and improve safety. Leveraging the edge computing paradigm, AI-based condition monitoring systems process vast streams of vehicular data to detect anomalies and optimize operational performance. 
In this work, we introduce a novel vehicle condition monitoring service that enables real-time diagnostics of a diverse set of anomalies while remaining practical for deployment in real-world edge environments. To address mobility challenges, we propose a closed-loop service orchestration framework where service migration across edge nodes is dynamically triggered by network-related metrics. Our approach has been implemented and tested in a real-world race circuit environment equipped with 5G network capabilities under diverse operational conditions. Experimental results demonstrate the effectiveness of our framework in ensuring low-latency AI inference and adaptive service placement, highlighting its potential for intelligent transportation and mobility applications.
\end{abstract}


\section{Introduction}
The integration of intelligent condition monitoring in vehicular systems is expected to play a pivotal role in the development of autonomous and connected mobility, ensuring reliability, sustainability, and efficiency in future transportation ecosystems. The staggering volume of vehicular measurement streams, driven by the widespread deployment of onboard sensors, in conjunction with expanded computational resources, offers enhanced monitoring capabilities and unlocks unprecedented application scenarios. As such, real-time condition monitoring of onboard vehicular equipment, fault provisioning, and predictive maintenance represent key services within the evolving vehicle-to-everything (V2X) paradigm. Building on massive data availability, such services leverage the proliferation of advanced artificial intelligence (AI) models and knowledge-extraction techniques to reveal hidden patterns, detect unknown correlations, and generate actionable intelligence with minimal human intervention. In turn, the timely detection of irregular patterns in monitoring information and the identification of anomalies/failures enhance operational efficiency, improve vehicular safety, and reduce unplanned downtime and maintenance costs. 

The ongoing modernization of vehicular systems heavily relies on seamless data acquisition and pervasive connectivity enabled by fifth-generation (5G) and emerging sixth-generation (6G) communication technologies. Besides meeting critical requirements in terms of ultra-low latency, high reliability, and node density, 5G- and 6G-based networks enhance the scalability and flexibility of vehicular systems through agile network deployments. In this context, emerging edge computing architectures aim at storing, processing, analyzing, and responding to data close to the acquisition sensors, enabling dramatically faster processing times and localized decision-making \cite{chen2023federated}. Locally fused and processed data at the edge combine received information from a set of onboard sensors to form a decentralized condition monitoring platform for supervising vehicular equipment indicators and proactive scheduling of maintenance operations. 
Nevertheless, edge-based vehicular diagnostics face challenges from variations in road conditions and frequent vehicle handovers, causing potential service disruption in high-mobility environments. 

A comprehensive review of AI technologies applied for vehicle maintenance is provided in \cite{Mahale2025}, highlighting their role in effective monitoring of health conditions.
The authors in \cite{9785863} proposed an
AI-based risk management system for remotely monitoring vehicular engine health status to reduce unplanned downtime. In a similar direction, 
a hybrid deep learning-based vehicular engine health monitoring system using convolutional neural networks (CNNs) was proposed in \cite{RAHIM2024125080}, demonstrating improved fault detection accuracy.
The authors in \cite{wang2023transformer} assessed the performance of transformer-based models in detecting rare fault patterns by capturing long-range dependencies in time-series data.
Nevertheless, traditional deep learning models, such as CNNs and transformer-based architectures, can be computationally intensive, hindering real-time decision making; thus, their applicability in edge environments with limited computational capabilities may be hindered.
The interplay of edge computing with federated learning was explored in \cite{chen2023federated} for decentralized fault diagnosis while preserving data privacy. Hybrid approaches combining physics-based models with AI-driven fault diagnostics have also shown promise in improving interpretability and robustness \cite{singh2021hybrid}. However, key challenges remain in handling diverse operational conditions with various anomaly patterns while ensuring low-latency inference.

A systematic review of service migration strategies in vehicular edge computing \cite{wang22} highlights the importance of proximity-based service placement in high-mobility vehicular environments. 
An edge-enabled V2X service placement strategy was proposed in \cite{8955944} to optimize resource utilization and improve service performance by strategically locating services at edge servers.
In \cite{9107503}, a joint service migration and mobility optimization approach using multi-agent deep reinforcement learning was introduced to minimize service migration delay in vehicular edge computing environments. However, traditional service migration strategies do not leverage real-time network event monitoring and typically operate using static or predictive models that rely on pre-defined handover thresholds or historical mobility patterns. Thus, such approaches may fail to adapt to sudden changes in network conditions, leading to service disruptions and inefficient resource utilization.

To address such challenges, this paper introduces a novel AI-based condition monitoring system for real-time vehicular anomaly detection at the edge, enhanced with closed-loop service orchestration where service migration is dynamically triggered by network-related metrics.
In particular, our contribution is threefold:
\begin{itemize}
    \item We develop a vehicular condition monitoring service that effectively monitors vehicular health while remaining deployable in real-world edge automotive environments by maintaining low-latency inference.  
    \item We introduce a service migration mechanism that ensures seamless transitions and service continuity during vehicle handovers by dynamically relocating resources based on real-time network events.
    \item We evaluate through real-world trials the practical feasibility and detection performance of our proposed framework, considering diverse anomaly patterns. 
\end{itemize}

The remainder of the paper is organized as follows. Section \ref{sec:three} presents the considered system model. Section \ref{sec:four} elaborates on the methodology followed for AI-based anomaly detection and service migration at the edge. Results pertaining to the detection performance of our monitoring service and the service migration are presented in Section \ref{sec:five}. Section \ref{sec:six} is reserved for the conclusions and discussion of the path forward.

\section{System Model} \label{sec:three}
The implementation phases for a vehicular condition monitoring service comprise: \textit{i}) identification of the critical assets in a vehicle, \textit{ii}) acquisition of monitoring information, \textit{iii}) data fusion, transmission, and processing, \textit{iv}) analysis of failure modes, and \textit{v}) decision-making. 

\begin{figure}[t!]
    \centering
    \includegraphics[width=1\columnwidth]{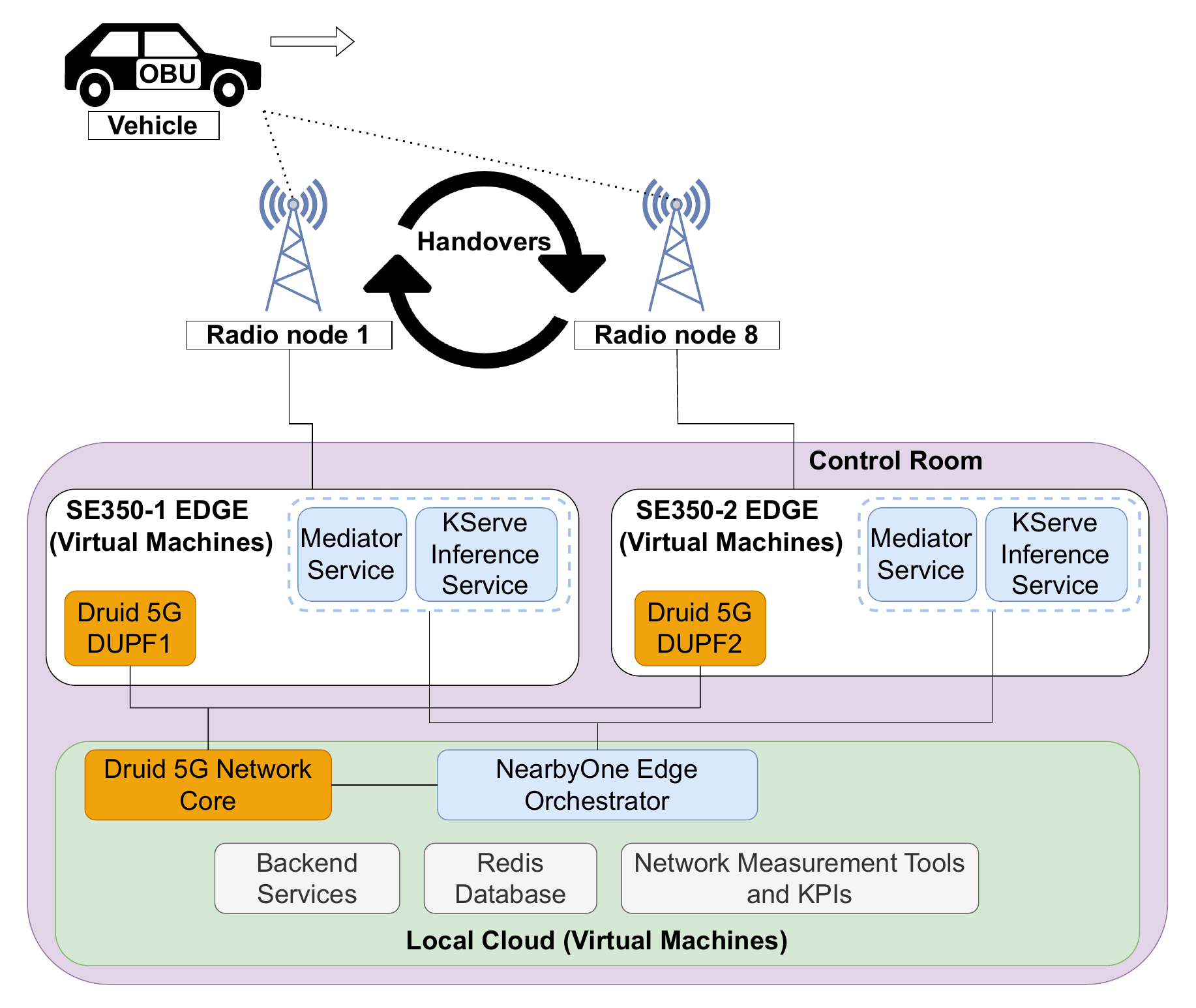}
    \caption{Edge infrastructure for vehicle condition monitoring system with C-V2X connectivity and associated services. Services related to the 5G standalone network are highlighted in orange, while all other services are relevant to the condition monitoring use case.}
    \vspace{-0.5cm}
    \label{fig:sysmodel}
\end{figure}

\subsection{Network Model}
Fig.~\ref{fig:sysmodel} illustrates the considered network scenario where an edge monitoring infrastructure is deployed to monitor the vehicle condition across its trajectory. The vehicle is equipped with monitoring sensors that acquire status information related to various vehicular operations. Transmission to the edge infrastructure takes place with the aid of a vehicular on-board unit (OBU), which fuses aggregated information from the sensors. Network connectivity is provided by 
cellular V2X (C-V2X) technology, offering scalable and seamless wireless communication between the vehicle and multiple radio nodes.
The transmitted measurement streams are then processed at the edge monitoring nodes, which offer computational resources close to the vehicle for real-time condition monitoring. 

A service orchestrator (in our case NearbyOne\footnote{https://www.nearbycomputing.com/wp-content/uploads/2021/09/Product-Data-Sheet-NBYCOMP.pdf}) is further considered to provide lifecycle management functionalities to the vehicular condition monitoring service at the edge. In the case of handovers, the service orchestrator performs service migration across the available edge nodes based on events originating from the 5G core network. 
Based on the knowledge extracted and with the help of appropriate visualization tools, instructive and actionable insights can be derived, e.g., flag whether a fault has occurred and diagnose its type in event-detection operations. Knowledge extraction may be further assisted by historical event logs and life expectancy statistics to predict when maintenance activities will be required.

\subsection{Supported Services}
Each edge node is associated with a radio node 
and comprises \textit{i}) a dedicated virtual machine for hosting the distributed user plane function (UPF) of its associated radio node; \textit{ii}) a virtual machine reserved for the condition monitoring service; and \textit{iii}) a third virtual machine, isolated from all others, used for the 5G network application programming interface (API) and network event calls. Specifically, the condition monitoring service involves two microservices: \textit{i}) a Mediator service which adapts the OBU messages into the appropriate inference request format; and \textit{ii}) a KServe inference service enabling scalable and low-latency real-time inference directly from vehicle measurement streams.

The local cloud provides a cluster of virtual machines that support different services. The 5G core network is deployed within a dedicated virtual machine, isolated from all other services. Another isolated virtual machine hosts all backend components, including Grafana,
a visualization tool that creates dashboards to monitor system performance and model behavior. The Redis database for storing vehicular data has been deployed in a third virtual machine, while a fourth virtual machine is allocated for network measurement tools. 

\section{Methodology and Implementation}\label{sec:four}
\subsection{AI-based Condition Monitoring Service}
\subsubsection{Machine Learning Model for Anomaly Detection}
In this work, we employ a light gradient boosting machine (LightGBM) classification model for anomaly detection, leveraging its advanced features in processing large volumes of sequential sensor readings and detecting subtle patterns indicative of potential equipment failures. By training on historical sensor data, LightGBM can effectively classify different types of faults and determine their root causes. Its superior capability to detect anomalies with limited training data has been demonstrated in our previous work~\cite{9947986}, where it was evaluated against deep learning models, including TabTransformer~\cite{huang2020tabtransformer} and TabNet~\cite{arik2020tabnet}, which served as benchmark classifiers.

To efficiently optimize hyperparameters in our LightGBM-based classifier, we leverage Bayesian optimization techniques using the Optuna library~\cite{optuna_2019}. Unlike traditional methods such as grid search or random search, which explore the hyperparameter space in a brute-force or purely stochastic manner, Bayesian optimization 
builds a probabilistic model to predict promising or informative hyperparameter values, improving search efficiency over traditional grid/random search. As such, the search process is guided effectively based on prior evaluations, significantly reducing the computational cost.
Additionally, we incorporate a pruning mechanism to facilitate the early stopping of unpromising trials based on intermediate validation performance. This further enhances efficiency by allocating computational resources to more promising configurations.

\subsubsection{Model Deployment at the Edge}
The LightGBM model was deployed at each edge node using KServe, enabling scalable and low-latency inference directly from vehicle telemetry streams. Measurements were collected and stored in InfluxDB, allowing precise and systematic analysis of inference behavior.


\subsection{Edge-Orchestrated Service Migration}
\subsubsection{Handover Process and Edge Node Transitions}
During vehicle handovers, the vehicle
changes its connection between tracking area identities (TAIs) and the condition monitoring service running at the edge must migrate to ensure service continuity.
Radio nodes within a TAI utilize a distributed UPF for local breakout to an edge server, enabling efficient data handling and reduced latency. The UPF selection, based on the radio node's TAI, ensures association with the geographically closest UPF. Consequently, when a vehicle moves between TAIs, the service must migrate to the new edge server across the vehicle's trajectory to maintain continuity.

\subsubsection{Service Orchestration Mechanism}
The service orchestrator manages service migrations by monitoring vehicle mobility. In particular, it subscribes to \texttt{smf\_sm\_context} events from the 5G core's session management function (SMF) to track the vehicle's current TAI during active data sessions. Upon detecting a TAI change, indicating a handover, the 5G core notifies the orchestrator through these events. Leveraging its knowledge of both the edge monitoring infrastructure and network (i.e., via SMF events), the orchestrator determines the appropriate UPF where the vehicle's user plane will be directed to and when a service migration is necessary. Upon receiving an event, the orchestrator processes it and checks whether the monitoring service is running on the edge node connected through the UPF to the new TAI. If the service is not running on the appropriate edge server, the orchestrator initiates service migration to ensure that the service follows the vehicle's trajectory and is available at the new edge node.

\subsubsection{Service Migration Strategy}
To ensure seamless and efficient transitions of the monitoring service between edge locations, minimizing downtime and performance degradation, the service migration strategy employs the following steps upon receiving an event indicating the need for migration:
\begin{enumerate}
    \item \underline{Event Processing}: The orchestrator extracts relevant details (e.g., PLMN ID, TAI) from the event and maps them to the corresponding edge node.
    \item \underline{Service Status Verification}: The current state of the service is verified to determine if migration is required.
    \item \underline{Service Relocation}: If the service runs on a different edge node, the orchestrator updates its configuration to migrate it to the correct edge node.   
    \item \underline{Polling and Confirmation}: The orchestrator polls the service status at regular intervals (every 0.5s) until the status changes to \texttt{SERVICESTATUS\_IN\_SYNC}, ensuring that the migration has been successfully completed.   
    \item \underline{Migration Time Calculation}: The time elapsed for migration is computed by logging the timestamp of the migration request and subtracting it from the timestamp when \texttt{SERVICESTATUS\_IN\_SYNC} is verified. This data is stored in a persistent volume for future analysis. Migration time includes high-level orchestration time, which is the time the orchestrator takes to perform the action, and low-level orchestration time, which is the time the service needs to run in the edge node's resource orchestration platform (e.g., Kubernetes).
\end{enumerate}

\section{Performance Evaluation} \label{sec:five}
\subsection{Experimental Setup}
\begin{figure}[t!]
    \centering
    \subfloat[OBU used in trials]{\includegraphics[width=0.47\columnwidth]{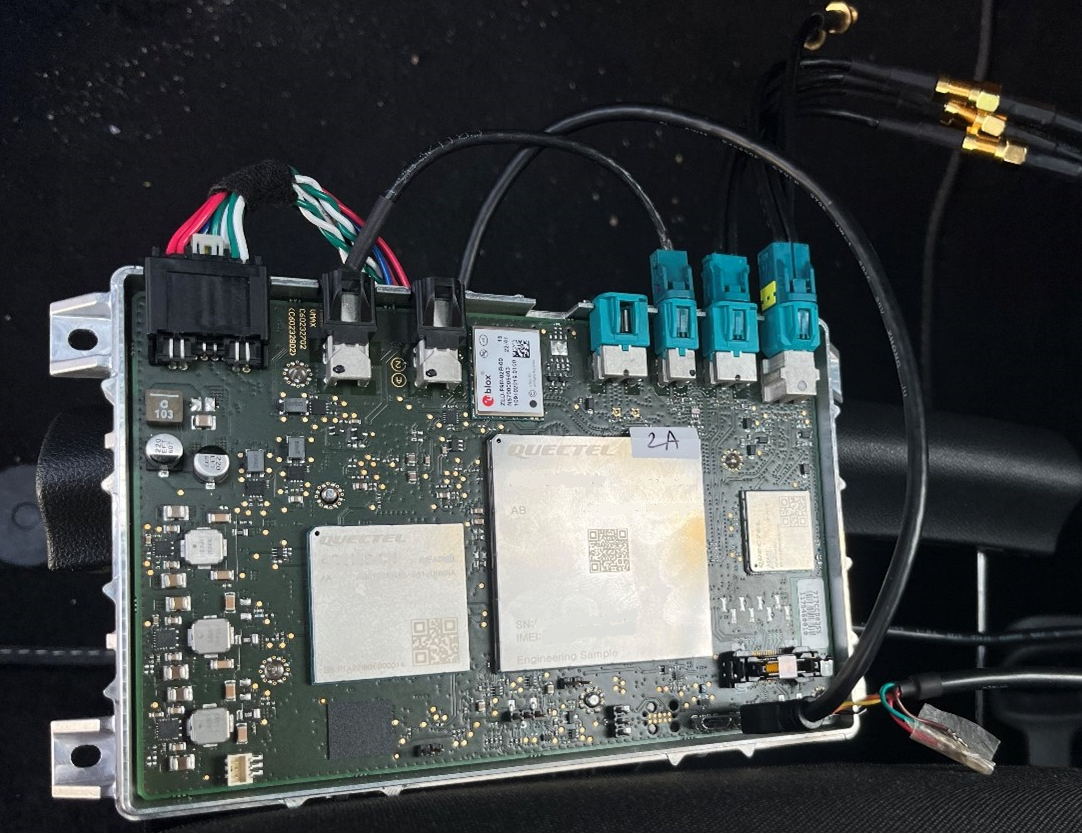}}
    \hfill 
    \subfloat[Radio node]{\includegraphics[width=0.47\columnwidth]{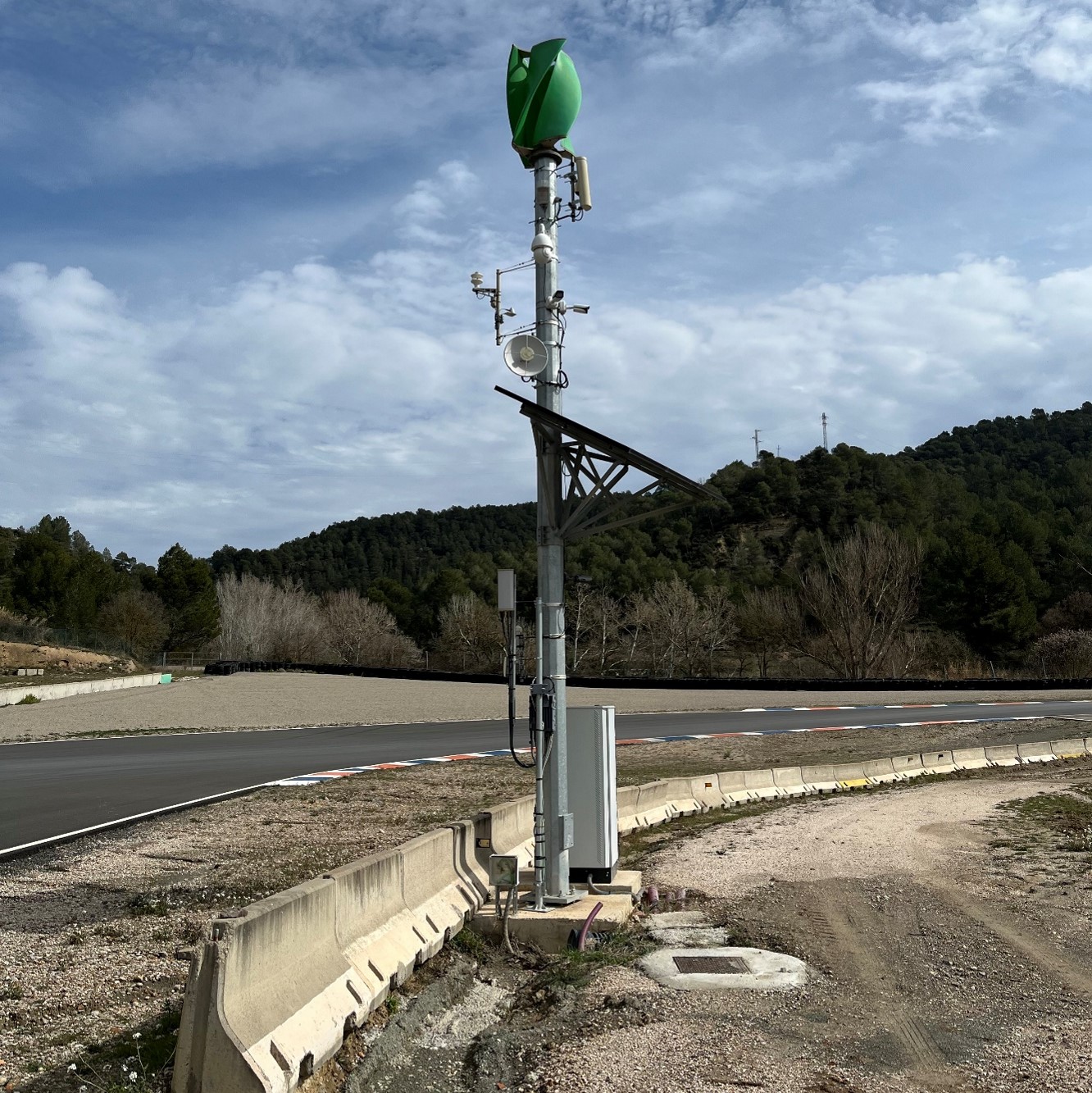}} \\
    \vspace{-0.2cm}
    \subfloat[Edge servers in the control room]{\includegraphics[width=0.47\columnwidth]{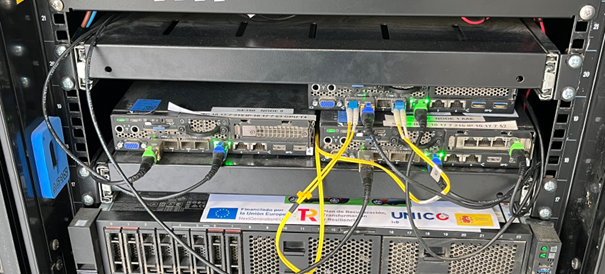}}
    \hfill
    \subfloat[Vehicle approaching radio node]{\includegraphics[width=0.47\columnwidth]{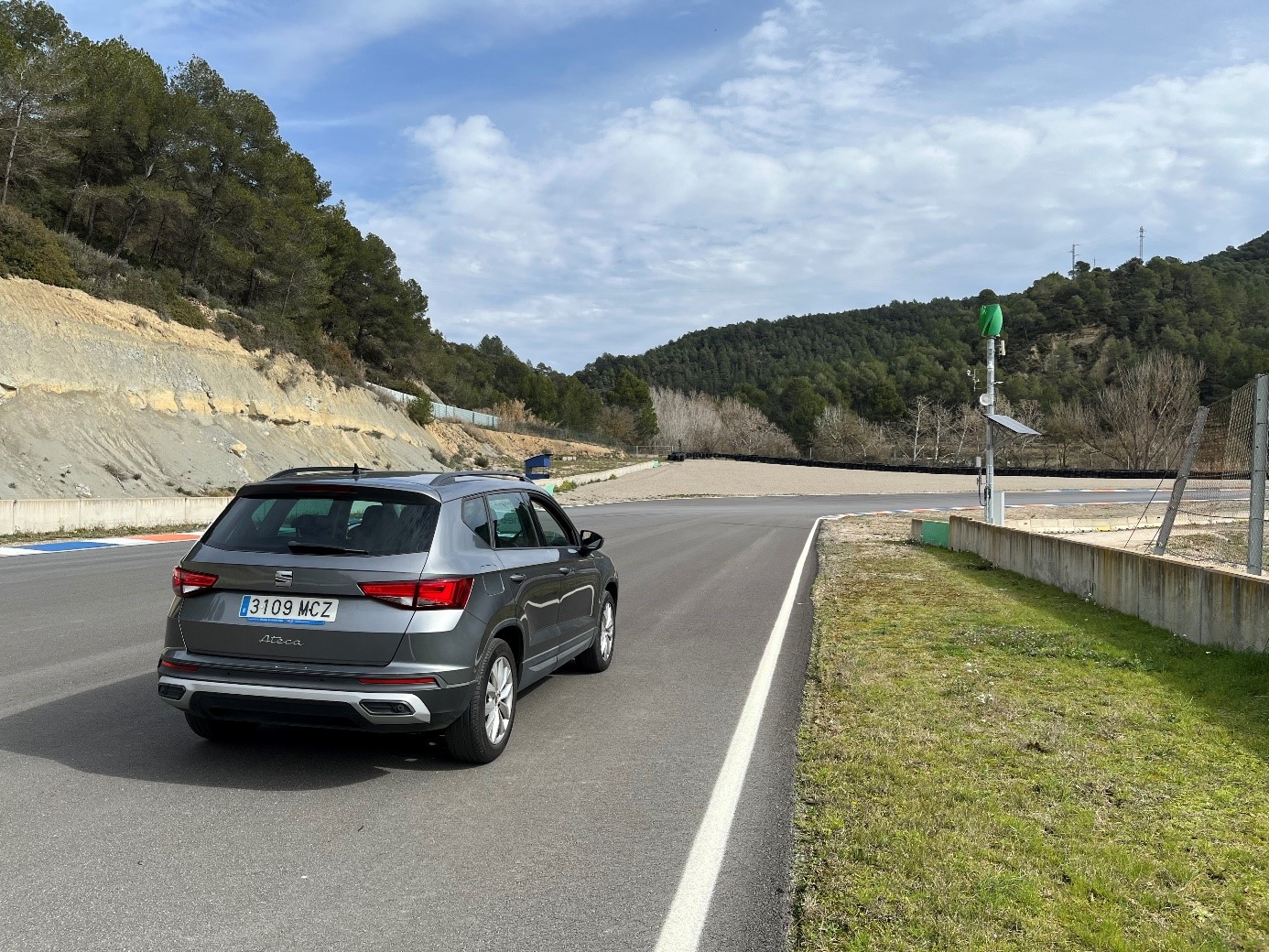}\label{fig:seat}}
    \caption{Experimental setup at the Circuit Parcmotor Castellolí.}
    \label{fig:four_subfigures}
    \vspace{-0.5cm}
\end{figure}

A series of real-world tests were carried out at the Cellnex Mobility Lab at Circuit Parcmotor Castellolí to assess the performance of our condition monitoring service and the integration of its enabling components. Fig. \ref{fig:four_subfigures} depicts various elements of the monitoring infrastructure, as outlined in Fig. \ref{fig:sysmodel}. A SEAT Ateca R4 2.0 TDI vehicle, equipped with monitoring sensors, continuously transmitted operational status data via its C-V2X OBU to the edge monitoring infrastructure with the aid of two radio nodes providing connectivity. The 5G Standalone mobile network utilized the Raemis™ Druid software solution, a 3GPP-compliant 5G core with REST API and additional functionalities. The edge computing infrastructure 
comprised two servers, each associated with a radio node, hosting virtual machines for the dockerized services (Section \ref{sec:three}). 
The vehicle's trajectory involved multiple handovers between the radio nodes and their respective edge servers. Service orchestration was managed by the NearbyOne orchestrator.

\begin{figure}[t!]
    \centering
    \includegraphics[width=\columnwidth]{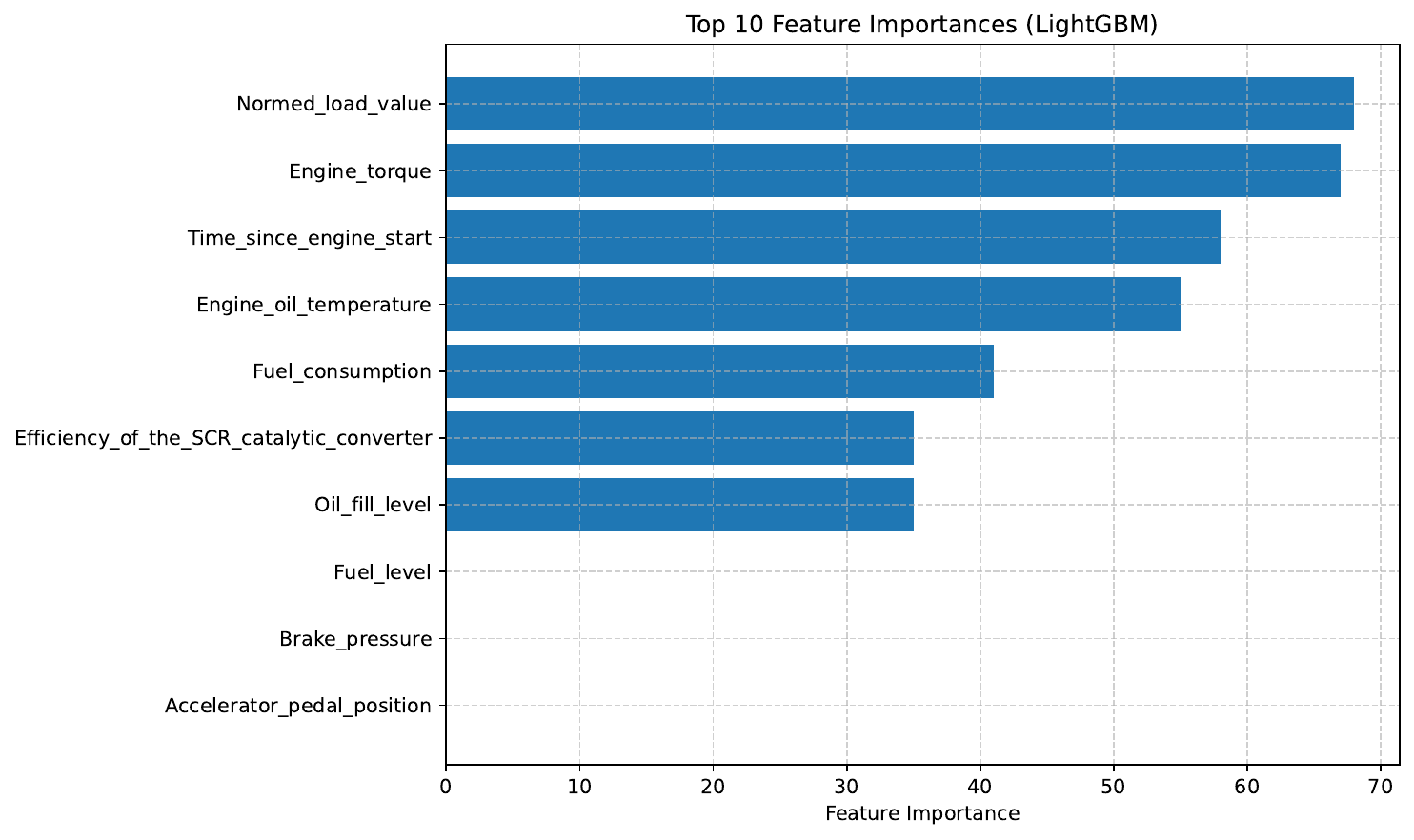}
    \caption{Feature importance analysis for anomaly detection.}
    \vspace{-0.35cm}
    \label{fig: feat}
\end{figure}
\begin{figure}[t!]
    \centering
    \includegraphics[width=\columnwidth]{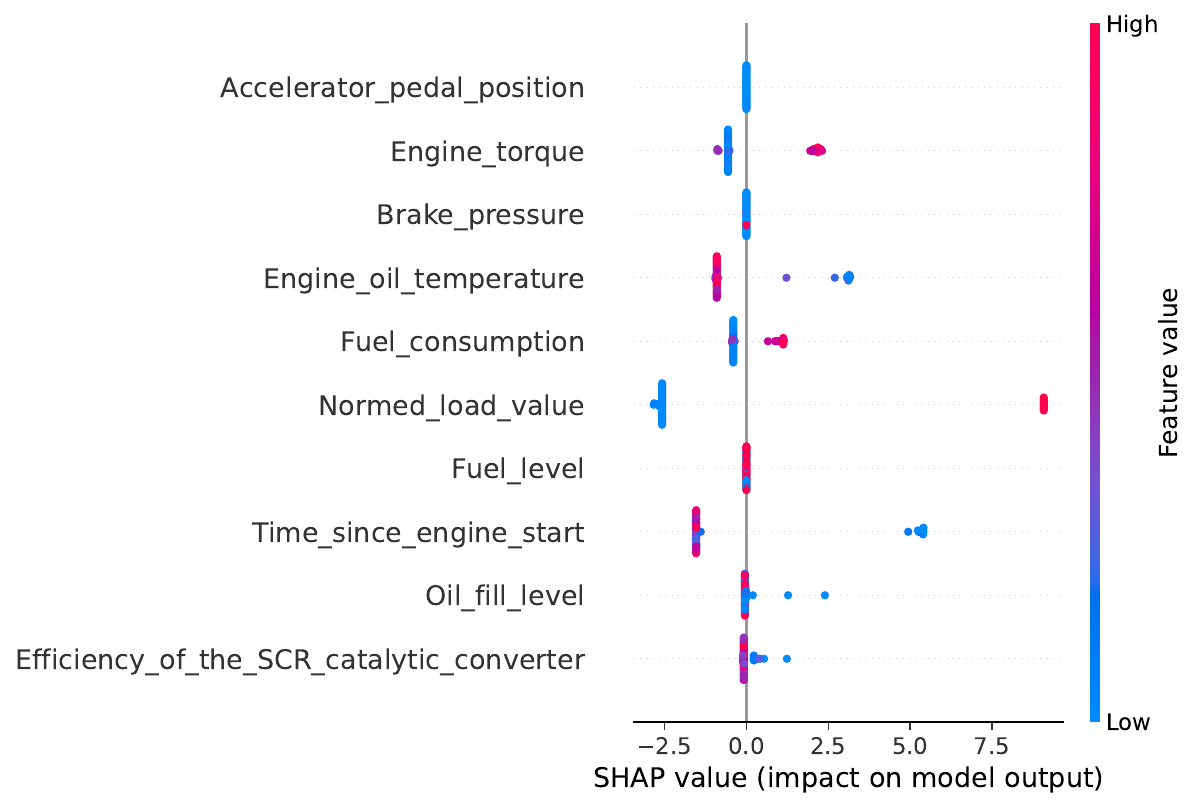}
    \caption{SHAP value impact on LightGBM model outputs.}
    \label{fig: shap}
    \vspace{-0.5cm}
\end{figure}

\subsubsection{Model Training}
For model training, real-world telemetry data from the vehicle were considered, covering diverse driving scenarios (urban, highway, and idle conditions). After loading, datasets were concatenated, cleaned, and preprocessed to remove constant columns, unify formats, and standardize variable names. A synthetic anomaly class (0/1) was assigned to simulate anomaly detection scenarios. To ensure model transparency and validate the learned patterns, a feature importance analysis was conducted using the trained LightGBM model. Fig. \ref{fig: feat} depicts the key features influencing anomaly detection. It can be observed that the model focuses strongly on variables linked to engine load, torque, time, and critical performance sensors, while other variables (e.g., fuel level, brake pressure, accelerator pedal position) had no impact on the final decision-making.
To add an additional layer of explainability, SHAP (SHapley Additive exPlanations) values were computed to analyze individual and global feature contributions of the model outputs. SHAP is a widely used explainability technique that leverages cooperative game theory to quantify the impact of each feature on the model’s classification decisions \cite{NIPS2017_7062}. The SHAP summary plot in Fig. \ref{fig: shap} offers insights into how features influence predictions toward normal or anomalous classes.

\subsubsection{Model Testing}
\begin{figure}[t!]
    \centering
    \includegraphics[width=\columnwidth]{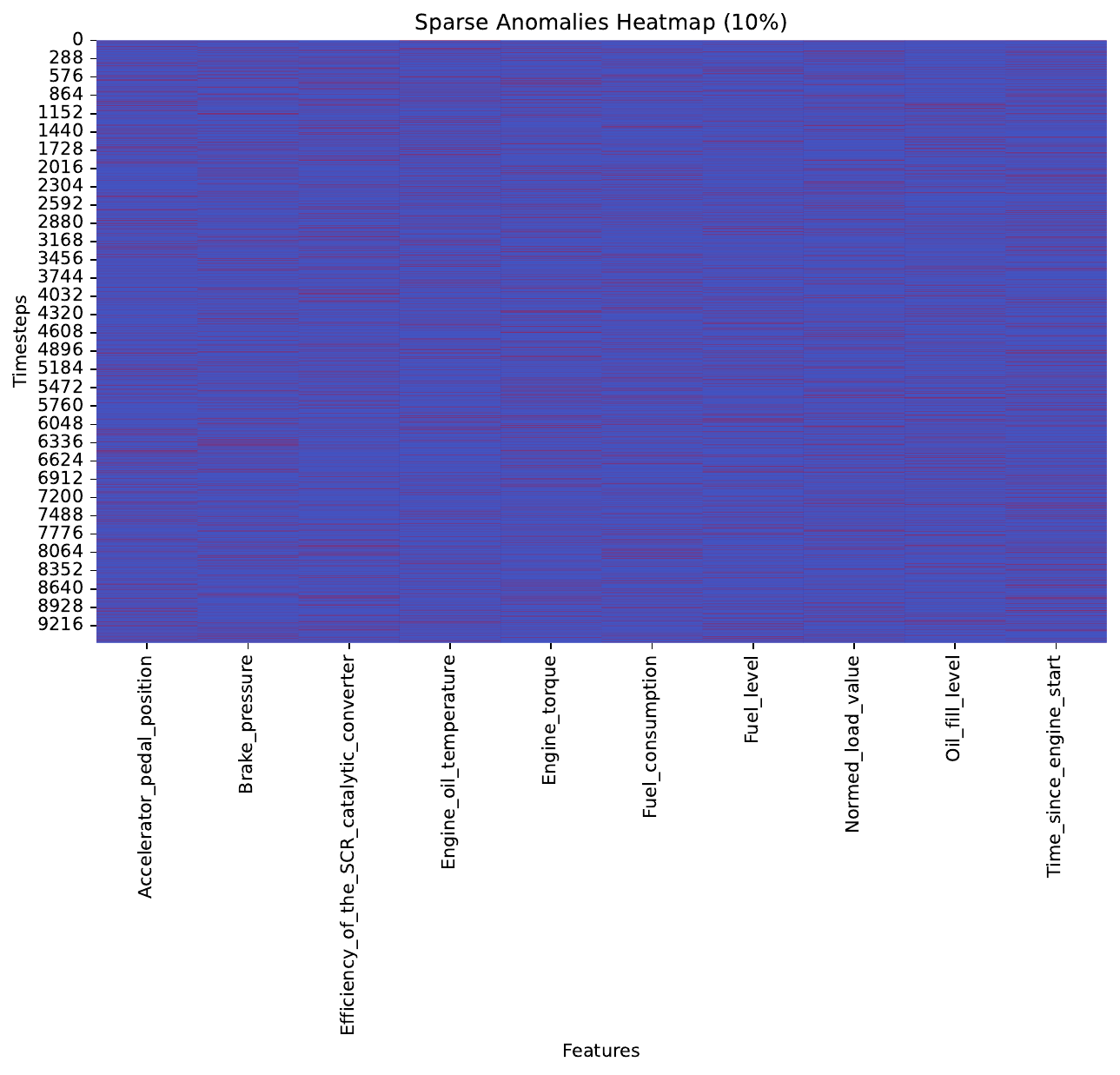}
    \caption{Sparse anomalies heatmap where synthetic faults are introduced in 10\% of the dataset.}
    \label{fig:sparse}
    \vspace{-0.2cm}
\end{figure}

\begin{figure}[t!]
    \centering
    \includegraphics[width=\columnwidth]{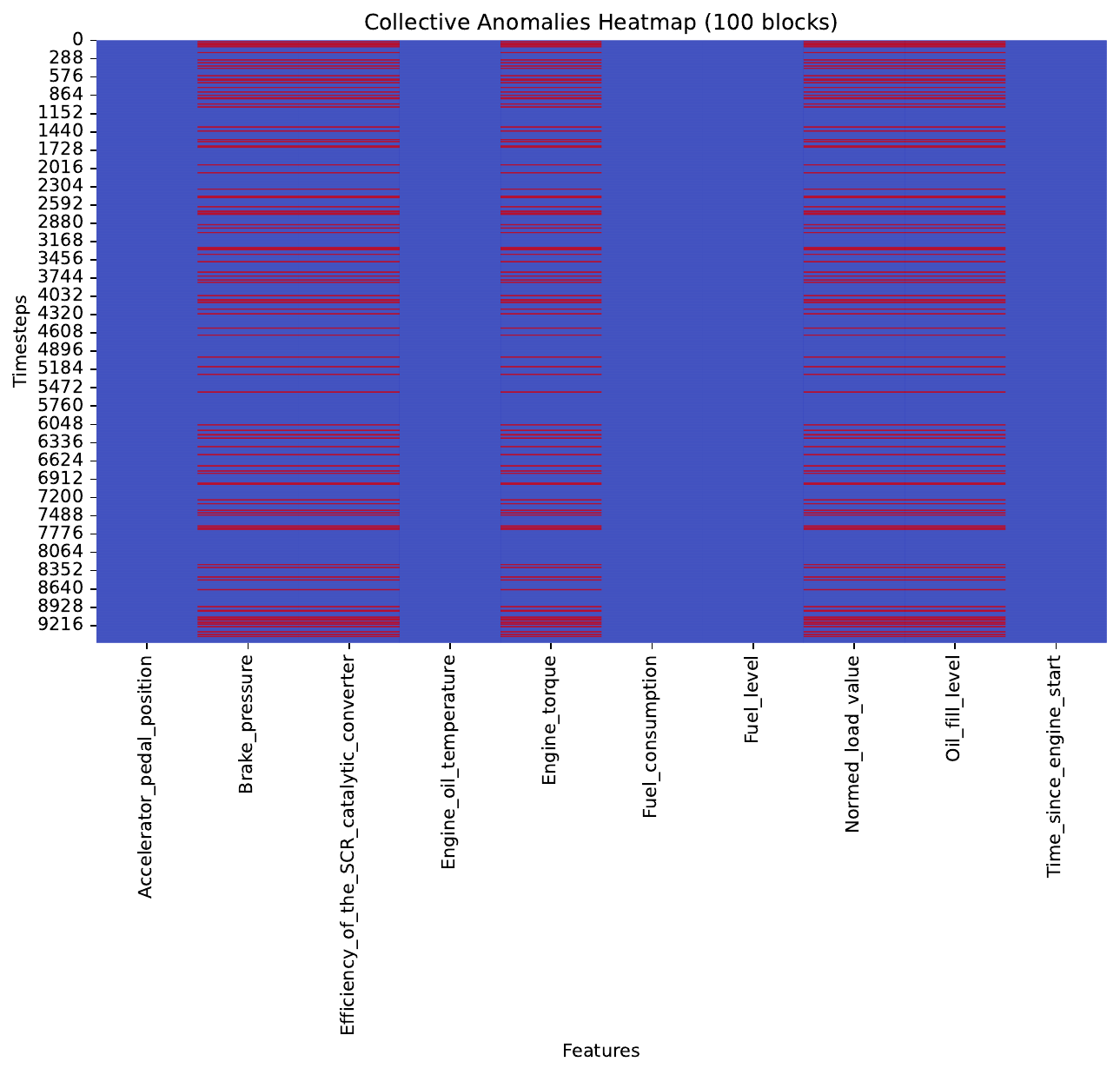}
    \caption{Collective anomalies heatmap for 100 consecutive timesteps.}
    \vspace{-0.2cm}
    \label{fig:collective}
\end{figure}

The testing dataset contains 9,487 telemetry records and retains the same features as the training set, ensuring consistency for evaluation. Based on the statistical analysis shown in Table \ref{table:summary_statistics},
we introduced synthetic anomaly values across the different features, as in Table \ref{table:anomaly_injection}.
\begin{table}[t!]
    \centering
    \caption{Overview of the distribution and spread of key features in the testing dataset}
    \begin{adjustbox}{max width=\columnwidth}
    \begin{tabular}{p{0.34\columnwidth}ccccccc}
        \hline
        \textbf{Feature} & \textbf{Mean} & \textbf{Std} & \textbf{Min} & \textbf{25\%} & \textbf{50\%} & \textbf{75\%} & \textbf{Max} \\
        \hline
        Accelerator pedal position & 16.93 & 9.62 & 0.00 & 14.90 & 14.90 & 14.90 & 86.30 \\
        Brake pressure & 1.61 & 29.29 & 0.00 & 0.00 & 0.00 & 0.00 & 655.33 \\
        Efficiency of the SCR catalytic converter & \multirow{2}{*}{0.76} & \multirow{2}{*}{0.25} & \multirow{2}{*}{0.00} & \multirow{2}{*}{0.64} & \multirow{2}{*}{0.88} & \multirow{2}{*}{0.92} & \multirow{2}{*}{0.98} \\
        Engine oil temperature & 9.65 & 28.59 & 0.00 & 0.00 & 0.00 & 0.00 & 113.20 \\
        Engine torque & 38.75 & 45.51 & 0.00 & 25.80 & 28.60 & 32.40 & 392.90 \\
        Fuel consumption & 0.69 & 3.45 & 0.00 & 0.00 & 0.00 & 0.00 & 29 \\
        Fuel level & 2.47 & 7.58 & 0.00 & 0.00 & 0.00 & 0.00 & 28 \\
        Normed load value & 29.68 & 15.80 & 0.00 & 23.80 & 25.70 & 28.70 & 97.70 \\
        Oil fill level & 70.23 & 10.35 & 0.00 & 67.23 & 71.51 & 73.93 & 92.09 \\
        Time since engine start & 470.17 & 1634.49 & 0.00 & 0.00 & 0.00 & 0.00 & 10190 \\
        \hline
    \end{tabular}
    \end{adjustbox}
    \vspace{-0.1cm}
    \label{table:summary_statistics}
\end{table}
\begin{table}[t!]
    \centering
    \caption{Anomaly injection values}
    \vspace{-0.1cm}
    \begin{tabular}{lc}
        \hline
        \textbf{Feature} & \textbf{Anomalous Value} \\ 
        \hline
        Accelerator pedal position & 200 \\
Brake pressure & 2000 \\
Efficiency of the SCR catalytic converter & 2.0 \\
Engine oil temperature & 250 \\
Engine torque & 700 \\
Fuel consumption & 70 \\
Fuel level & 100 \\
Normed load value & 200 \\
Oil fill level & 200 \\
Time since engine start & -100 \\
        \hline
    \end{tabular}
    \label{table:anomaly_injection}
    \vspace{-0.3cm}
\end{table}
The following anomaly patterns are considered:
\begin{itemize}
    \item \textit{Sparse} anomalies (Fig. \ref{fig:sparse}): Random isolated anomalies were injected into 1\%, 5\%, and 10\% of the dataset to simulate faults across scattered features and timesteps.
    \item \textit{Collective} anomalies (Fig. \ref{fig:collective}): Sequential temporal anomalies were injected using a varying (i.e., 10, 100, and 200) number of consecutive timesteps. In this case, half of the features were selected randomly for anomaly injection, creating partial system failures.
   \end{itemize}

\subsection{Anomaly Detection Performance}
\begin{figure}[t!]
    \centering
    \includegraphics[width=\columnwidth]{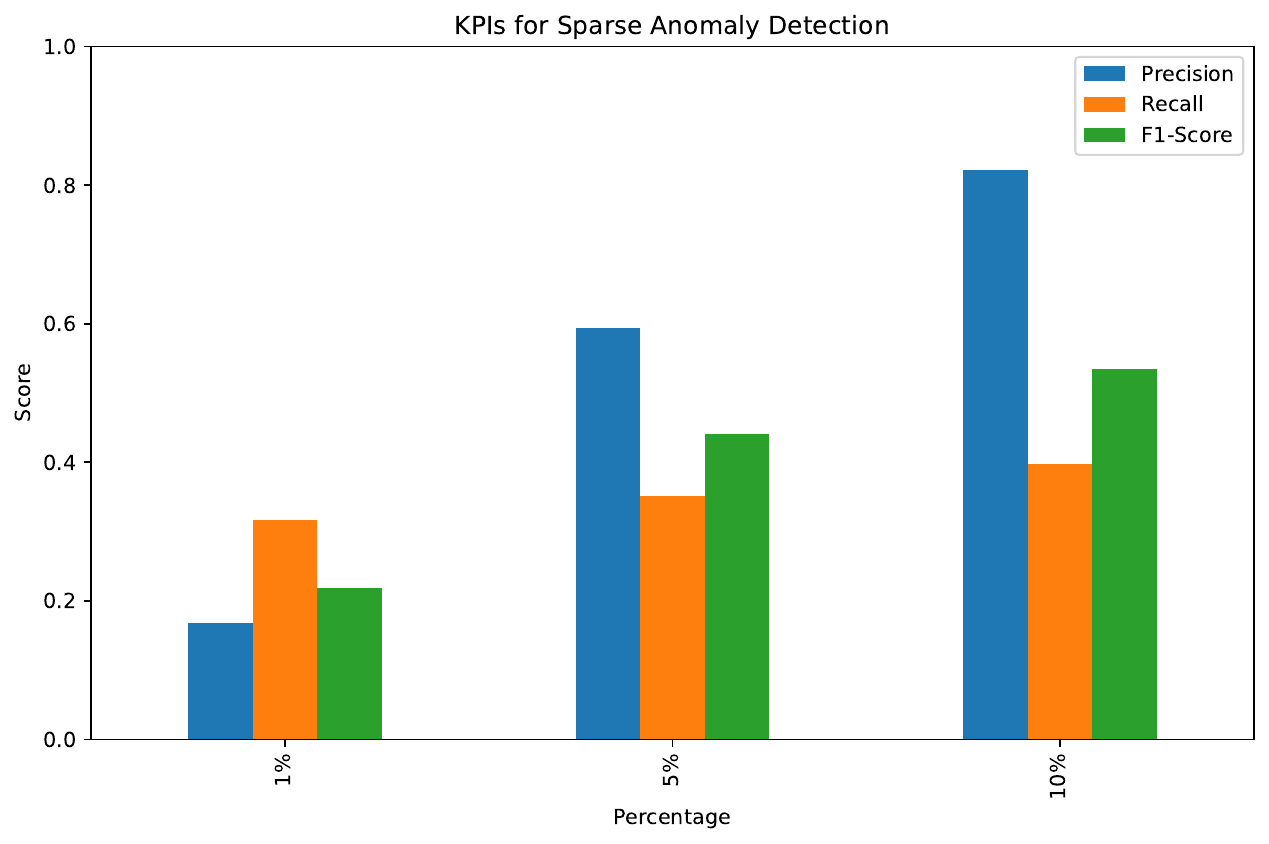}
    \caption{Detection performance for a varying percentage of sparse anomalies.}
    \label{fig:det-sparse}
\end{figure}

\begin{figure}[t!]
    \centering
    \includegraphics[width=\columnwidth]{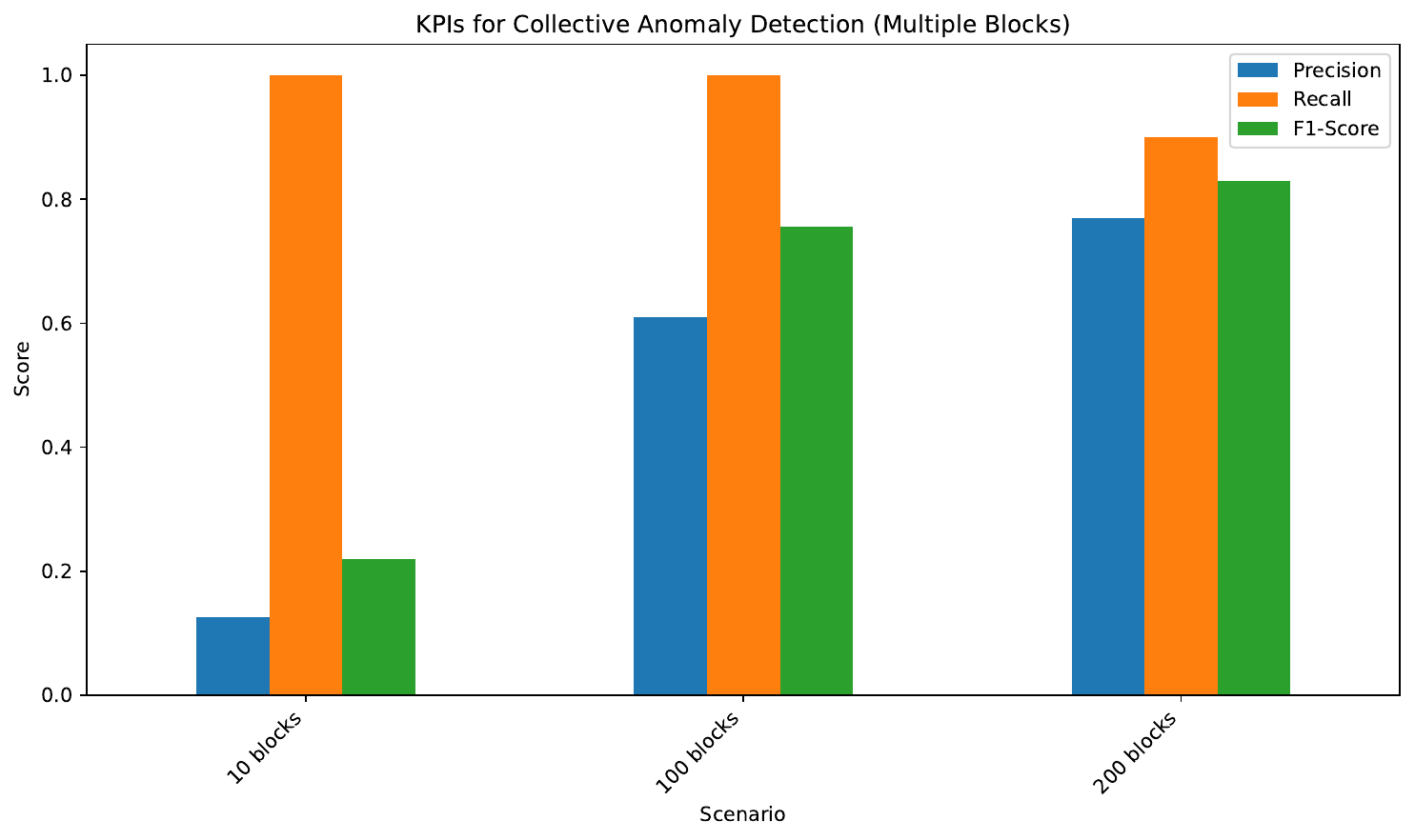}
    \caption{Detection performance for a varying number of consecutive timesteps with collective anomalies.}
    \label{fig:det-collective}
\end{figure}

The performance of our LightGBM-based classifier for the detection of sparse and collective anomalies is illustrated in Figs. \ref{fig:det-sparse} and \ref{fig:det-collective}, respectively. By jointly evaluating widely used metrics, i.e., precision, recall, and F1-score, we aim to provide insights into the robustness and effectiveness of fault detection. Precision and recall values capture the impact of false-positive and false-negative rates, respectively. The F1-score represents the harmonic mean between precision and recall metrics.

In the case of sparse anomalies (Fig. \ref{fig:det-sparse}), precision improves remarkably as anomaly density increases, revealing that false positives become less frequent when anomalies occur more often. However, recall remains consistently low across all anomaly densities, suggesting that the model struggles to detect many actual anomalies, leading to a high false negative rate. Although the F1-score improves with increasing anomaly density, it remains at moderate levels. As such, the model could benefit from techniques aimed at enhancing its sensitivity to sparse faults \cite{10817499}.

In the case of collective anomalies (Fig. \ref{fig:det-collective}), recall values of 1.0 for 10 and 100 consecutive timesteps indicate that these anomalies can be accurately detected with zero false negatives. Notably, recall remains at high levels even for 200 consecutive timesteps, confirming that the model reliably detects collective anomalies affecting contiguous time windows.
Precision improves as the number of timesteps increases, demonstrating that the model becomes more accurate in distinguishing real anomalies when they are more frequent, as seen also in sparse anomaly scenarios. Finally, the F1-score shows significant growth as collective anomalies become more evident, revealing enhanced detection performance in identifying sustained equipment faults.

In addition to evaluating model detection performance, we also monitored the average time during model inference deployed via KServe. In particular, as shown in Table \ref{table:time-summary}, an average model inference time of 15.1ms was reported, ensuring real-time responsiveness of the condition monitoring service, which is a critical requirement for diagnostic applications in connected mobility environments.

\subsection{Service Migration Time}
Service migration time was evaluated in two experimental conditions: with and without the inclusion of the Mediator microservice (Section \ref{sec:three}). The Mediator, a Redis consumer that adapts OBU messages for inference and forwards responses to the observability backend, was observed to impact migration duration, as shown in Table \ref{table:time-summary}. The experiments revealed that the mean migration time with the Mediator microservice was approximately 66.75s, with a standard deviation of 10.93s. Without the mediator, the mean migration time decreased significantly to approximately 24.57s, with a standard deviation of 3.39s. It is worth noting that these measurements provide valuable practical insights into the deployment and applicability of AI tools within the system. More specifically, they indicate the required prediction horizon to ensure that services are deployed on the new edge node in a timely manner, thereby guaranteeing uninterrupted service continuity, supporting the real-time condition monitoring. 

\begin{table}[t!]
    \centering
    \caption{Inference and service migration times}
    \vspace{-0.1cm}
    \begin{adjustbox}{max width=\columnwidth}
    \begin{tabular}{p{0.5\columnwidth}cc}
        \hline
        \textbf{Time} & \textbf{Mean Value (s)} & \textbf{Std (s)} \\
        \hline
        KServe inference time & 0.0151 & 0.002\\
        Migration time (w/ Mediator service) & 66.754& 10.926\\
        Migration time (w/o Mediator service) & 24.57 & 3.39\\
        \hline
    \end{tabular}
    \end{adjustbox}
    \label{table:time-summary}
    \vspace{-0.5cm}
\end{table}

\section{Conclusion}\label{sec:six}
\vspace{-0.08cm}
In this paper, a novel AI-based vehicular condition monitoring system was introduced for real-time anomaly detection at the edge, enhanced with closed-loop service orchestration. Real-world trials were conducted to assess the practical feasibility and detection performance of our framework under diverse operational conditions. Our approach was shown to achieve low-latency inference time and service continuity across the edge infrastructure.
While the detection of collective anomalies in contiguous time windows was effectively performed, the efficacy in detecting isolated anomalies was limited. Enhancing sensitivity to sparse faults is therefore a key focus of our future work.
\vspace{-0.13cm}


\bibliographystyle{IEEEtran}
\bibliography{refs.bib}

\begin{thebibliography}{10}
\providecommand{\url}[1]{#1}
\csname url@samestyle\endcsname
\providecommand{\newblock}{\relax}
\providecommand{\bibinfo}[2]{#2}
\providecommand{\BIBentrySTDinterwordspacing}{\spaceskip=0pt\relax}
\providecommand{\BIBentryALTinterwordstretchfactor}{4}
\providecommand{\BIBentryALTinterwordspacing}{\spaceskip=\fontdimen2\font plus
\BIBentryALTinterwordstretchfactor\fontdimen3\font minus
  \fontdimen4\font\relax}
\providecommand{\BIBforeignlanguage}[2]{{%
\expandafter\ifx\csname l@#1\endcsname\relax
\typeout{** WARNING: IEEEtran.bst: No hyphenation pattern has been}%
\typeout{** loaded for the language `#1'. Using the pattern for}%
\typeout{** the default language instead.}%
\else
\language=\csname l@#1\endcsname
\fi
#2}}
\providecommand{\BIBdecl}{\relax}
\BIBdecl

\bibitem{chen2023federated}
L.~Chen, M.~Zhang, and R.~Gupta, ``{Federated Learning for Edge-Based Fault
  Diagnosis in Connected Vehicles},'' \emph{ACM Transactions on Cyber-Physical
  Systems}, vol.~7, no.~3, pp. 1--20, 2023.

\bibitem{Mahale2025}
Y.~Mahale \emph{et~al.}, ``{A comprehensive review on artificial intelligence
  driven predictive maintenance in vehicles: technologies, challenges and
  future research directions},'' \emph{Discover Applied Sciences}, vol.~7,
  2025.

\bibitem{9785863}
M.~A. Rahim \emph{et~al.}, ``{An Intelligent Risk Management Framework for
  Monitoring Vehicular Engine Health},'' \emph{IEEE Transactions on Green
  Communications and Networking}, vol.~6, no.~3, pp. 1298--1306, 2022.

\bibitem{RAHIM2024125080}
M.~A. Rahim, M.~M. Rahman, M.~S. Islam, A.~J.~M. Muzahid, M.~A. Rahman, and
  D.~Ramasamy, ``Deep learning-based vehicular engine health monitoring system
  utilising a hybrid convolutional neural network/bidirectional gated recurrent
  unit,'' \emph{{Expert Systems with Applications}}, vol. 257, p. 125080, 2024.

\bibitem{wang2023transformer}
T.~Wang, D.~Kim, and X.~Liu, ``{Transformer-Based Anomaly Detection for
  Automotive Sensor Networks},'' \emph{Neural Computing and Applications},
  vol.~35, no.~4, pp. 1125--1139, 2023.

\bibitem{singh2021hybrid}
R.~Singh, P.~Kumar, and A.~Bose, ``{Hybrid AI-Physics-Based Fault Diagnostics
  for Automotive Systems},'' \emph{Mechanical Systems and Signal Processing},
  vol. 160, p. 107905, 2021.

\bibitem{wang22}
L.~Wang and T.~Chen, ``A survey on service migration strategies for vehicular
  edge computing,'' in \emph{Lecture Notes in Computer Science}.\hskip 1em plus
  0.5em minus 0.4em\relax Springer, 2022.

\bibitem{8955944}
A.~Moubayed, A.~Shami, P.~Heidari, A.~Larabi, and R.~Brunner, ``{Edge-Enabled
  V2X Service Placement for Intelligent Transportation Systems},'' \emph{{IEEE
  Trans. on Mobile Computing}}, vol.~20, no.~4, pp. 1380--1392, 2021.

\bibitem{9107503}
Q.~Yuan \emph{et~al.}, ``{A Joint Service Migration and Mobility Optimization
  Approach for Vehicular Edge Computing},'' \emph{{IEEE Transactions on
  Vehicular Technology}}, vol.~69, no.~8, pp. 9041--9052, 2020.

\bibitem{9947986}
P.~Mulinka \emph{et~al.}, ``{Optimizing a Digital Twin for Fault Diagnosis in
  Grid Connected Inverters - A Bayesian Approach},'' in \emph{2022 IEEE Energy
  Conversion Congress and Exposition (ECCE)}, 2022, pp. 1--6.

\bibitem{huang2020tabtransformer}
X.~Huang \emph{et~al.}, ``Tabtransformer: Tabular data modeling using
  contextual embeddings,'' \emph{arXiv preprint arXiv:2012.06678}, 2020.

\bibitem{arik2020tabnet}
S.~O. Ar{\i}k and T.~Pfister, ``{TabNet: Attentive Interpretable Tabular
  Learning},'' \emph{arXiv preprint arXiv:1908.07442}, 2020.

\bibitem{optuna_2019}
T.~Akiba \emph{et~al.}, ``{Optuna: A Next-generation Hyperparameter
  Optimization Framework},'' in \emph{Proceedings of the 25rd {ACM} {SIGKDD}
  International Conference on Knowledge Discovery and Data Mining}, 2019.

\bibitem{NIPS2017_7062}
S.~M. Lundberg and S.-I. Lee, ``{A Unified Approach to Interpreting Model
  Predictions},'' in \emph{Advances in Neural Information Processing Systems
  30}, I.~Guyon, U.~V. Luxburg, S.~Bengio, H.~Wallach, R.~Fergus,
  S.~Vishwanathan, and R.~Garnett, Eds.\hskip 1em plus 0.5em minus 0.4em\relax
  Curran Associates, Inc., 2017, pp. 4765--4774.

\bibitem{10817499}
D.~Gutierrez-Rojas \emph{et~al.}, ``{Detection and Classification of Anomalies
  in WSN-Enabled Cyber-Physical Systems},'' \emph{IEEE Sensors Journal},
  vol.~25, no.~4, pp. 7193--7204, 2025.

\end{thebibliography}


\begin{thebibliography}{00}
\bibitem{b1} G. Eason, B. Noble, and I. N. Sneddon, ``On certain integrals of Lipschitz-Hankel type involving products of Bessel functions,'' Phil. Trans. Roy. Soc. London, vol. A247, pp. 529--551, April 1955.
\bibitem{b2} J. Clerk Maxwell, A Treatise on Electricity and Magnetism, 3rd ed., vol. 2. Oxford: Clarendon, 1892, pp.68--73.
\bibitem{b3} I. S. Jacobs and C. P. Bean, ``Fine particles, thin films and exchange anisotropy,'' in Magnetism, vol. III, G. T. Rado and H. Suhl, Eds. New York: Academic, 1963, pp. 271--350.
\bibitem{b4} K. Elissa, ``Title of paper if known,'' unpublished.
\bibitem{b5} R. Nicole, ``Title of paper with only first word capitalized,'' J. Name Stand. Abbrev., in press.
\bibitem{b6} Y. Yorozu, M. Hirano, K. Oka, and Y. Tagawa, ``Electron spectroscopy studies on magneto-optical media and plastic substrate interface,'' IEEE Transl. J. Magn. Japan, vol. 2, pp. 740--741, August 1987 [Digests 9th Annual Conf. Magnetics Japan, p. 301, 1982].
\bibitem{b7} M. Young, The Technical Writer's Handbook. Mill Valley, CA: University Science, 1989.
\end{thebibliography}

\end{document}